\documentclass[journal=apchd5,manuscript=article]{achemso}
 
\usepackage{amsmath,amssymb}
\usepackage{graphicx}
\usepackage{bm}

\makeatletter
\renewcommand*{\acs@author@list}{%
  Donghyun Park\textsuperscript{1,\,$\dagger$},
  Alex Song\textsuperscript{5},
  Haejun Chung\textsuperscript{1,\,2,\,3,\,*},
  and Sejeong Kim\textsuperscript{4,\,*}%
}

\renewcommand*{\acs@address@list}{\upshape%
  \textsuperscript{1}Department of Artificial Intelligence Semiconductor Engineering, Hanyang University, Seoul, Republic of Korea\par
  \textsuperscript{2}Department of Artificial Intelligence, Hanyang University, Seoul, Republic of Korea\par
  \textsuperscript{3}Department of Electronic Engineering, Hanyang University, Seoul, Republic of Korea\par
  \textsuperscript{4}Department of Electrical and Computer Engineering, Sungkyunkwan University, Suwon, Republic of Korea\par
  \textsuperscript{5}School of Electrical and Computer Engineering, The University of Sydney, Sydney, NSW, Australia\par
}
\renewcommand*{\acs@contact@details}{%
  \textsuperscript{*}Correspondence: haejun@hanyang.ac.kr; sejeongk@skku.edu\par
}
\acs@emailtrue
\makeatother
 
\author{Haejun Chung}
\affiliation{Hanyang University}
\author{Sejeong Kim}
\affiliation{Sungkyunkwan University}

\title{Inverse-Designed Metasurfaces for Compact Optical Skyrmion Generation with High Topological Fidelity}


\begin{document}


\begin{abstract}
Optical skyrmions are structured vector fields with nontrivial polarization topology and subwavelength-scale features. One common approach to generating optical skyrmions is the superposition of a zeroth-order Bessel beam and a higher-order Bessel beam carrying orbital angular momentum, with each beam possessing an orthogonal circular polarization state. However, creating such complex beams typically requires bulky free-space optical setups; therefore, recent efforts have focused on compact optical skyrmion generators based on metasurfaces. Nevertheless, achieving the degrees of freedom required for simultaneous phase and polarization control remains challenging because of the limited design flexibility of conventional meta-atoms. Here, we address this challenge by employing an inverse-design approach and demonstrate a single-layer metasurface that generates high-fidelity optical skyrmions. We employ an adjoint-based topology-optimization method to design a silicon metasurface that converts an incident beam into an optical skyrmion without the need for additional optical components. The optimized metasurface generates an optical skyrmion with skyrmion number $(N_\mathrm{sk}) = 0.970$. This work demonstrates that inverse design can be a promising route to compact skyrmion generators, and our approach provides a basis for near-field particle manipulation and the generation of independent topological bits in dense photonic integration. 
\end{abstract}

\section{Introduction}

Optical skyrmions are emerging topological quasiparticles of light that consist of spatially varying Stokes polarization vectors, and are characterized by their skyrmion number, which represents the number of times the polarization vector wraps around the Poincar\'{e} sphere.~\cite{Tsesses2018,Shen2022,Shen2024,teng2025construction,Ye2024} The skyrmion number remains resilient under continuous field deformations, provided that the relevant boundary conditions are preserved.~\cite{McWilliam2023,Ye2024,Wang2024Protection} This combination of topological robustness and vectorial structure has motivated the study of optical skyrmions for applications ranging from information encoding and high-precision metrology to robust data transfer through complex media, subwavelength imaging, super-resolution microscopy, and nanophotonic information processing.~\cite{Wang2025Arithmetic,Du2019,Wang2024Protection,He2024,Luo2023NonHermitian}

It was only recently that optical skyrmions were observed in optics, and so far there have been several demonstrations of methods for creating them.~\cite{Tsesses2018,Shen2024} Early demonstrations included deploying surface plasmons and their evanescent fields.~\cite{Tsesses2018} Other methods frequently use interferometers to coaxially superimpose two Bessel beams, in which the two Bessel beams have orthogonal circular polarization states and the higher-order Bessel beam carries orbital angular momentum (OAM).~\cite{Shen2022,Ye2024,Singh2023,SrinivasaRao2024} This requires bulky optical components such as phase retarders, polarizers, and spatial light modulators (SLMs) or digital micromirrors (DMDs) to impart OAM to one of the Bessel beams, which makes it challenging to miniaturize the system.~\cite{Shen2022,McWilliam2023,Shen2024} To reduce the size of the system, dielectric metasurfaces have been employed, revolutionizing the generation of optical skyrmions.~\cite{He2024,MataCervera2025} Meta-atoms are designed to precisely control the phase, amplitude, or polarization of light.~\cite{Arbabi2015,Overvig2019,so2023revisiting} So far, there are several simulation papers demonstrating metasurface-based generation of optical skyrmions, but there are fewer experimental demonstrations.~\cite{MataCervera2025,He2024} Although metasurfaces have reduced the size and increased compactness, a quarter-wave plate (QWP) is still often required at the input/output beam path to encode circular polarization onto linear beams.~\cite{Shen2022,He2024,MataCervera2025} This is because creating a complex structural beam like an optical skyrmion is challenging, as it requires independent, simultaneous manipulation of polarization, amplitude, and phase. In short, the degrees of freedom required to create a Bessel-beam-based optical skyrmion exceed the capability of a standard meta-atom.~\cite{Arbabi2015,Overvig2019,Li2022review} This challenge has been addressed by various approaches, including using double-layer metasurfaces or exploiting structural birefringence and geometric phase to add degrees of freedom in the design space.~\cite{Mirzapourbeinekalaye2022,MataCervera2025}

In this work, we report the first inverse-designed metasurface for optical skyrmion generation. Adjoint optimization is used to obtain a single-layer metasurface that converts a normally incident $45^{\circ}$-linearly polarized Gaussian beam with beam-waist radius $w_0=1.15~\mu\mathrm{m}$ into an optical skyrmion field formed by the coherent superposition of a zeroth-order Bessel constituent (right-circularly polarized, zero topological charge) and a first-order Bessel constituent (left-circularly polarized, topological charge of one). In this simulation methodology, we incorporate spatial confinement directly into the target field to reduce the size of the optical skyrmion. For this, we use a full-wave inverse design that optimizes the entire aperture as a coupled system and imposes the optical objective directly on the observation plane, which can produce complex vector fields with high fidelity.~\cite{Li2022review,White2023} This formulation is naturally suited to skyrmion generation, where the target is a vector field specified by its amplitude, phase, and polarization at a single observation plane.~\cite{lin2022end,pestourie2018inverse,piggott2015inverse,LalauKeraly2013,yang2020inverse,Molesky2018,Fan2020,chung2020tunable,kim2025inverse} We truncate the two Bessel beams at their respective first zeros to include only the inner rings in the simulation, and demonstrate the formation of the optical skyrmion. As a result, we achieve a 6.00~$\mu$m-diameter binary silicon metasurface which, when optimized by adjoint-based topology optimization, reproduces the prescribed skyrmion field with a field overlap of 96.0\% at $\lambda = 1550$~nm, yields a simulated skyrmion number of $N_\mathrm{sk}=0.970$, and limits the cumulative leakage outside the region of interest to below 5.00\%, compared with approximately 50.0\% for the corresponding unconfined skyrmion over the same radial extension. The optimized structure produces a compact optical skyrmion with high topological fidelity and strongly reduced peripheral leakage.
\begin{figure*}[!ht]
  \centering
  \includegraphics[width=\textwidth]{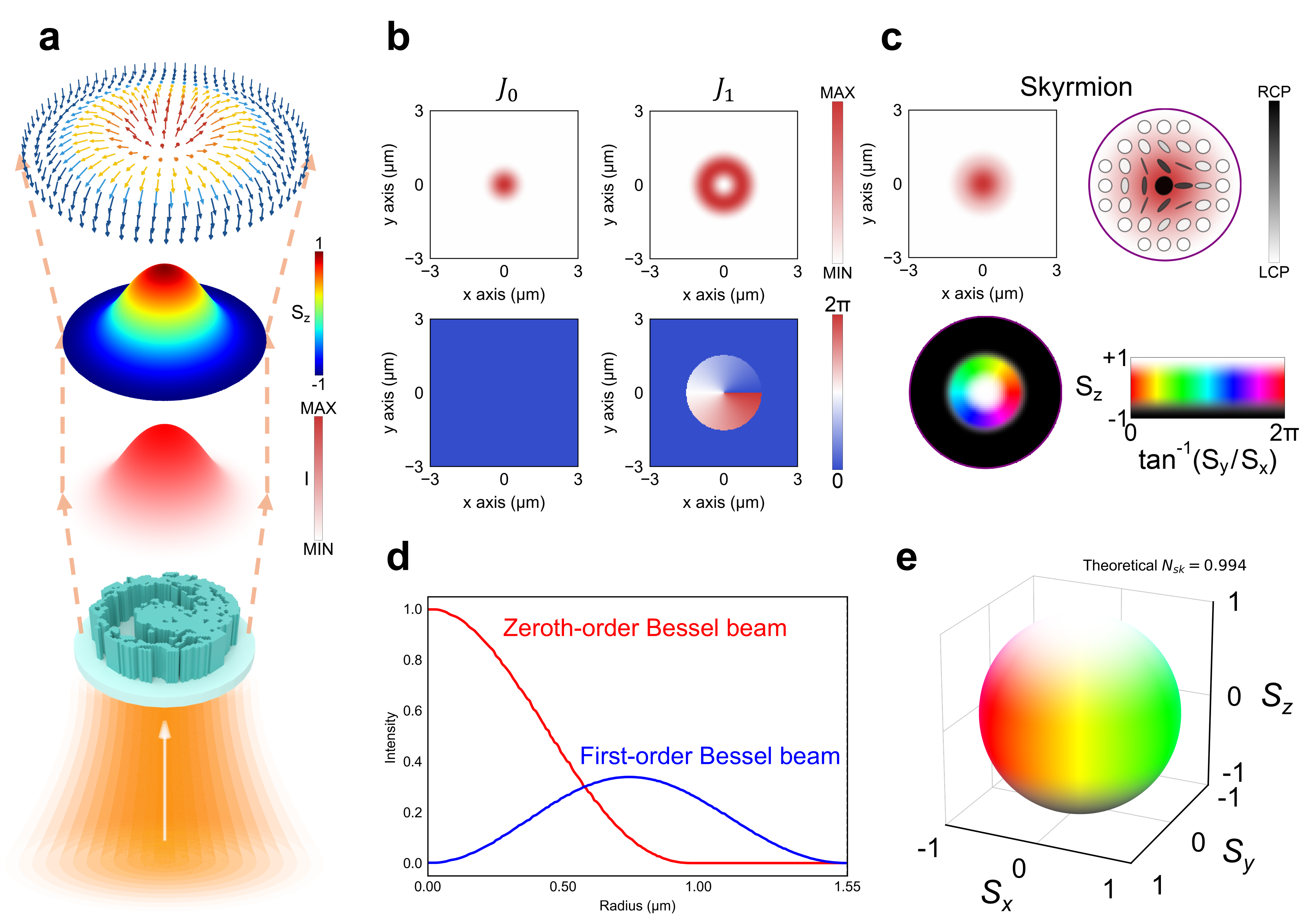}
  \caption{Inverse-designed metasurface for optical skyrmion generation. (a) Schematic illustration of the inverse-designed silicon metasurface. A single-layer metasurface converts a normally incident $45^{\circ}$-linearly polarized Gaussian beam into an optical skyrmion. From top, the desired polarization-vector field, the normalized out-of-plane Stokes parameter $(S_z)$, and the intensity profile $(I)$. (b) The desired field is composed of two orthogonal vector channels; the zeroth-order Bessel beam $(J_0)$ and the first-order Bessel beam $(J_1)$. Their corresponding amplitudes (top) and phase distributions (bottom) are shown, respectively. Here, both Bessel beams include only the inner ring in their intensity profiles, while the outer concentric rings are neglected; in this work, we refer to this as a truncated Bessel desired beam. (c) The superposition of $J_0$ and $J_1$ components creates the resulting desired skyrmion field. The intensity profile, the polarization-ellipse map, and the skyrmion texture are presented. In the accompanying skyrmion-texture color bar, the vertical axis denotes the normalized out-of-plane Stokes parameter $S_z$, and the horizontal axis denotes the Stokes-space azimuthal angle, written here as $\tan^{-1}(S_y/S_x)$. (d) Radial intensity profiles of the RCP $J_0$ and LCP $J_1$ components, illustrating truncation at their first radial zeros. (e) Mapping of the desired polarization field onto the Poincar\'{e} sphere, yielding a theoretical skyrmion number $(N_\mathrm{sk}) = 0.994$.}
  \label{fig:desired_field}
\end{figure*}

\section{Desired Field and Inverse-Design Framework}
The central task is to synthesize a spatially isolated optical skyrmion at a single observation plane. Doing so requires more than prescribing a local phase profile: the metasurface must reproduce the amplitude, phase, and polarization structure of the desired field while suppressing the extended outer rings associated with Bessel-mode skyrmions. We therefore first define the desired field, then describe the full-wave optimization workflow, and finally introduce the overlap-centered objective, adjoint formulation, and numerical setting used to realize it. Throughout this section, the observation plane is the common plane on which the desired field is prescribed, the figure of merit is evaluated, and the reported field quantities are extracted.

\subsection{Desired Field Construction and Device Geometry}
To formulate the inverse-design problem, we first define the desired field on a single observation plane and then specify the physical domain. Throughout this manuscript, we refer to this single plane simply as the observation plane. It is the plane on which the desired field is prescribed, the objective is evaluated in FDTD, and all reported field plots and metrics are extracted. The desired field is constructed to retain the essential information needed to create an optical skyrmion, which includes amplitude, phase, and polarization, while remaining compatible with a practical dielectric metasurface implementation. The device geometry, material platform, illumination condition, and observation plane are then fixed to define the boundary conditions and fabrication-aware constraints used throughout the optimization and validation steps.

Figure~\ref{fig:desired_field} summarizes the desired field and physical configuration considered in this work. As illustrated in Fig.~\ref{fig:desired_field}(a), the device is a disk-shaped silicon design region with a diameter of 6.00~$\mu$m and a thickness of 0.800~$\mu$m on a silica substrate. A normally incident Gaussian beam illuminates the structure from the substrate side, and the transmitted field is sampled at an observation plane located $z_\mathrm{obs}=1.55~\mu$m above the metasurface. In all forward simulations, the input source is a Gaussian beam whose waist is centered on the metasurface plane, with beam-waist radius $w_0=1.15~\mu\mathrm{m}$ and $45^{\circ}$ linear polarization with respect to the $x$- and $y$-axes. This is the only plane at which the desired field is defined and the output field is evaluated; in the FDTD implementation, the objective is evaluated and the adjoint source is constructed on this same observation plane. The design objective is to generate a compact vector field whose polarization texture forms a skyrmion within a finite transverse disk. The selected silicon-on-silica platform, together with the imposed minimum-feature and minimum-gap constraints introduced below, is compatible with established CMOS-compatible dielectric-metasurface fabrication routes such as electron-beam lithography, deep-ultraviolet lithography, and nanoimprint lithography~\cite{Arbabi2015,kamali2018review,Colburn2018,Moon2024LPR}.

At the observation plane, we introduce cylindrical coordinates $r=\sqrt{x^2+y^2}$ and the real-space azimuthal angle $\phi$, written here as $\tan^{-1}(y/x)$ with the usual quadrant convention, together with the transverse wavevector $k_\rho=\mathrm{NA}_{\mathrm{des}}\,k_0$, where $k_0=2\pi/\lambda$ and $\mathrm{NA}_{\mathrm{des}}=0.61$. For $\lambda=1550$~nm, the desired field is constructed as the coherent superposition of a right-circularly polarized zeroth-order Bessel constituent and a left-circularly polarized first-order vortex-Bessel constituent,
\begin{equation}
  \mathbf{E}_{\mathrm{des}}(r,\phi)
  = J_0(k_\rho r)\,\Theta(r_0-r)\,\hat{\mathbf e}_{\mathrm{R}}
  + J_1(k_\rho r)e^{i\phi}\,\Theta(r_1-r)\,\hat{\mathbf e}_{\mathrm{L}},
  \label{eq:desired_field}
\end{equation}
where $\Theta$ denotes the Heaviside step function, and the circular-basis unit vectors are defined by $\hat{\mathbf e}_{\mathrm{R}}=(\hat{\mathbf x}-i\hat{\mathbf y})/\sqrt{2}$ and $\hat{\mathbf e}_{\mathrm{L}}=(\hat{\mathbf x}+i\hat{\mathbf y})/\sqrt{2}$. For numerical optimization, Eq.~\ref{eq:desired_field} is resolved into Cartesian desired-field components $E_{\mathrm{des},x}$ and $E_{\mathrm{des},y}$ because the FDTD solver stores the observation-plane fields as $E_x$, $E_y$, and $E_z$. Circular-basis quantities are then reconstructed from $E_x$ and $E_y$ only when handedness-sensitive metrics are evaluated. The truncation radii are set by the first zeros of the corresponding Bessel functions,
\begin{equation}
  r_0=\alpha_{0,1}/k_\rho, \qquad
  r_1=\alpha_{1,1}/k_\rho,
  \label{eq:radii}
\end{equation}
with $\alpha_{0,1}=2.40$ and $\alpha_{1,1}=3.83$, yielding $r_0\approx0.97~\mu\mathrm{m}$ and $r_1\approx1.55~\mu\mathrm{m}$. The outer truncation radius $r_1$ also defines the circular desired-field region of interest, $A=\{(x,y): r\le r_1\}$, used later in the overlap, energy-concentration, and skyrmion-number calculations. The full observation-plane monitor is a larger circular window, $\Omega=\{(x,y):r\le r_\mathrm{m}\}$ with $r_\mathrm{m}=3.00~\mu\mathrm{m}$, so the ROI occupies only the central part of the monitored field. Unless otherwise noted, the circular field maps shown later are cropped to $A$, whereas the leakage analysis uses the entire monitor window $\Omega$.

Writing the desired field in terms of a right-circularly polarized $J_0$ term and a left-circularly polarized $J_1e^{i\phi}$ term makes the role of the two constituent fields explicit. The RCP $J_0$ term provides the central lobe with nearly uniform phase, whereas the LCP $J_1e^{i\phi}$ term supplies the annular envelope together with the azimuthal phase winding required for the skyrmionic polarization texture. Because $J_0(r_0)=0$ and $J_1(r_1)=0$, truncation at these order-specific first zeros keeps the field amplitude continuous at the truncation boundaries while removing the outer Bessel rings that would otherwise broaden the transverse footprint.

The annular region $A_\mathrm{ring}=\{(x,y):r_0<r\le r_1\}$ is especially important: in the ideal desired field, the RCP $J_0$ contribution has already vanished there, so this annulus should be dominated by the LCP $J_1e^{i\phi}$ constituent. This constituent separation later motivates a separate annular handedness term in the FoM, because residual RCP content in $A_\mathrm{ring}$ acts as noise that distorts the skyrmion boundary and degrades the skyrmion-number evaluation over $A$.

The resulting coherent superposition produces the compact desired skyrmion field shown in Fig.~\ref{fig:desired_field}(c), including the localized intensity distribution, a polarization-ellipse map plotted on the total-intensity background of the superposed skyrmion field, with ellipse grayscale indicating local handedness, and the skyrmion texture. In this skyrmion-texture representation, the accompanying two-dimensional color bar encodes the normalized out-of-plane Stokes parameter $S_z$ along the vertical direction and the Stokes-space azimuthal angle, written here as $\tan^{-1}(S_y/S_x)$, along the horizontal direction. Optically, $S_z$ specifies the local circular-polarization content and therefore the ellipticity and handedness of the polarization state: $S_z=\pm1$ correspond to the two circular-polarization poles on the Poincar\'e sphere, whereas $S_z=0$ lies on the linear-polarization equator. Written here as $\tan^{-1}(S_y/S_x)$ with the usual quadrant convention, this Stokes-space azimuthal angle gives the azimuth of the normalized Stokes-vector projection in the $S_x$-$S_y$ plane, i.e., the longitude around the Poincar\'e sphere. The skyrmion-texture color bar can therefore be interpreted as a two-dimensional encoding of the same Stokes-vector information that is mapped onto the Poincar\'e sphere in Fig.~\ref{fig:desired_field}(e). Figure~\ref{fig:desired_field}(d) further shows the radial intensity profiles of the RCP $J_0$ and LCP $J_1$ constituents, illustrating truncation of the $J_0$ component at its first zero $r_0$ and of the $J_1$ component at its first zero $r_1$. Finally, the Poincar\'e-sphere mapping in Fig.~\ref{fig:desired_field}(e) covers nearly the full solid angle and yields a theoretical skyrmion number of $N_\mathrm{sk}=0.994$, confirming that the truncated construction preserves the essential topology of the desired field while maintaining strong spatial localization.

\subsection{Observation-Plane Inverse-Design Workflow}

\begin{figure*}[!ht]
  \centering
  \includegraphics[width=\textwidth]{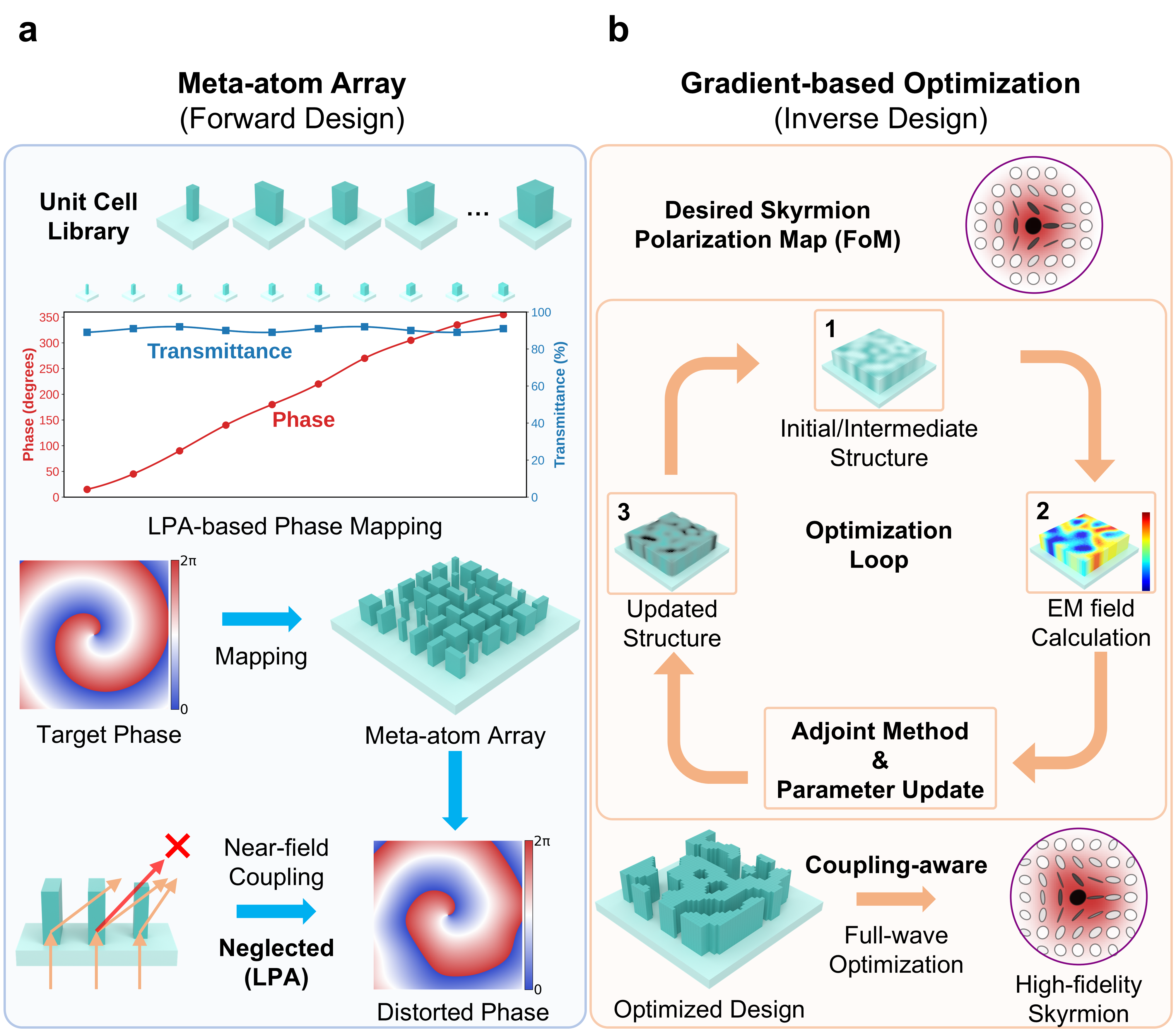}
  \caption{Comparison between a conventional forward-design workflow and the inverse-design workflow. (a) Schematic illustration describing a metasurface design strategy for controlling the phase of the incoming beam. A unit-cell library, which consists of various meta-atoms with different structural dimensions, is first constructed, and their phases are calculated. Once the desired phase is set, meta-atoms are distributed over two-dimensional space to impart the phase change to the incoming beam. However, this method misses effects arising from nearby structures and coupling. Therefore, the final phase becomes distorted compared with the intended phase. (b) Schematic illustration describing the gradient-based inverse-design framework. The desired skyrmion field is defined directly on the observation plane as the objective, and the three-dimensional structure is updated via repeated electromagnetic simulations, adjoint sensitivity analysis, and parameter updates so that the emitted field reproduces the desired field itself.}
  \label{fig:workflow}
\end{figure*}

Figure~\ref{fig:workflow} contrasts a conventional library-based design route with the present inverse-design strategy. In the conventional LPA-based forward workflow considered here, the optical specification is first reduced to a desired local phase profile and then approximated by assigning meta-atoms from a precomputed unit-cell library. More generally, library-based design can also encode local amplitude or polarization responses, but its logic remains one of local-response mapping. For the present problem, by contrast, the specified quantity is the desired skyrmion field itself: a compact vector field with coupled amplitude, azimuthal phase winding, polarization handedness, and a finite boundary across the full aperture.

The present workflow instead adopts an observation-plane inverse-design strategy in which the desired field itself is prescribed and matched directly through full-wave simulation and adjoint updates~\cite{White2023,Deng2026}. As shown in Fig.~\ref{fig:workflow}(b), the optimization starts from an initial three-dimensional structure, performs a full-wave electromagnetic simulation, evaluates a differentiable figure of merit (FoM) from the simulated field on the observation plane, computes the gradient with the adjoint method, and updates the geometry in a closed loop. Because the entire structure is re-simulated at every iteration, inter-element coupling is included throughout the optimization. The design goal is therefore not to realize a local library map, but to make the emitted vector field on the observation plane match the desired skyrmion field as closely as possible.

\subsection{Observation-Plane Objective and Numerical Setting}

\begin{figure*}[!ht]
  \centering
  \includegraphics[width=\textwidth]{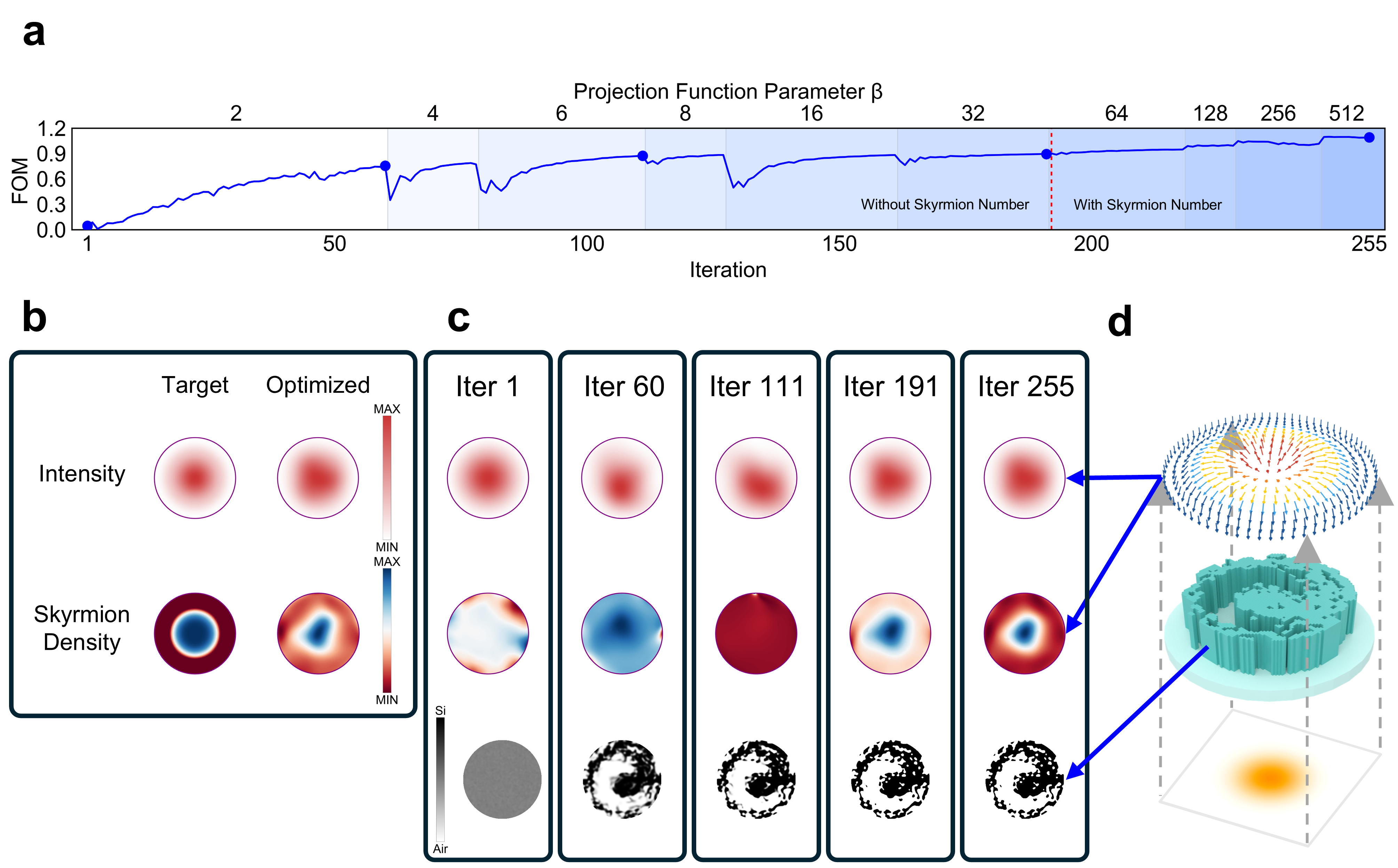}
  \caption{Inverse design of the optical skyrmion. (a) Figure of merit (FoM) as a function of iteration during the optimization. The projection-function parameter $\beta$ is progressively increased to enforce binarization of the design, and the skyrmion-number score is activated in the later stage of the optimization, as indicated by the red dashed line. (b) Comparison between the desired and optimized observation-plane fields within the ROI $A=\{(x,y):r\le r_1\}$, showing the intensity distribution (top) and the skyrmion-density map (bottom). (c) Representative intermediate states at selected iterations, again cropped to $A$, including the intensity profile (top), skyrmion-density distribution (middle), and corresponding material layout in the design region (bottom). The optimization gradually evolves from the initial structure toward a binary silicon--air pattern that more closely reproduces the desired field. (d) Schematic of the final optimized device, in which a normally incident $45^{\circ}$-linearly polarized Gaussian beam is transformed by the disk-shaped design region into the desired skyrmionic vector field above the metasurface.}
  \label{fig:optimization}
\end{figure*}

The optimization geometry and monitored quantities are summarized in Fig.~\ref{fig:optimization}(d). A cylindrical silicon design region is optimized so that the transmitted vector field on the observation plane reproduces the compact desired skyrmion field within the circular ROI $A$. In contrast to local phase-mapping approaches, the present formulation directly optimizes the full transmitted vector field, thereby enabling simultaneous control of field overlap, spatial confinement, and polarization topology. The core optical objective is therefore the agreement between the simulated and desired transverse fields on that plane, while additional terms concentrate total energy inside $A$, suppress residual RCP contamination in the annulus $A_\mathrm{ring}$, and guide the final topology.

In the discussion below, we use the term \emph{fidelity} to refer to agreement between the simulated output field and the prescribed desired field on the observation plane, as quantified here through the overlap, confinement, and topology metrics introduced below.

The field-level similarity between the simulated and desired fields is quantified by the normalized overlap,
\begin{equation}
  \eta
  = \frac{\left|\displaystyle\iint_A
    \mathbf{E}_\mathrm{sim}(x,y)\cdot
    \mathbf{E}_\mathrm{des}^{\ast}(x,y)\,dA\right|^2}
  {\left(\displaystyle\iint_A |\mathbf{E}_\mathrm{sim}(x,y)|^2\,dA\right)
   \left(\displaystyle\iint_A |\mathbf{E}_\mathrm{des}(x,y)|^2\,dA\right)},
  \label{eq:overlap}
\end{equation}
where $\mathbf{E}_\mathrm{sim}=(E_x,E_y)$ and $\mathbf{E}_\mathrm{des}=(E_{\mathrm{des},x},E_{\mathrm{des},y})$ denote the transverse vector fields on the observation plane, and $A=\{(x,y):r\le r_1\}$ is the circular desired-field region. Only the transverse components enter Eq.~\ref{eq:overlap} because the desired skyrmion field is prescribed in the transverse circular basis. This overlap is the primary beam-forming term in the optimization because it measures how closely the emitted vector field matches the desired field at the observation plane.

To quantify spatial isolation during optimization, we formulate the confinement-control term directly as the fraction of total monitor-plane energy concentrated inside the ROI,
\begin{equation}
  C_\mathrm{ROI}
  =
  \frac{\displaystyle\iint_A \left(|E_x|^2+|E_y|^2+|E_z|^2\right)\,dA}
       {\displaystyle\iint_\Omega \left(|E_x|^2+|E_y|^2+|E_z|^2\right)\,dA},
  \label{eq:roi_concentration}
\end{equation}
where $\Omega=\{(x,y):r\le r_\mathrm{m}\}$ is the full circular observation-plane monitor with $r_\mathrm{m}=3.00~\mu\mathrm{m}$. The term includes $E_z$ so that the confinement score reflects the total field energy relative to the full monitor, independent of the polarization basis used to define the desired skyrmion. Writing the FoM in the form of Eq.~\ref{eq:roi_concentration}, rather than as $1$ minus an outside leakage, makes the intended design logic explicit: even when the entire monitor window is considered, nearly all of the transmitted energy should remain inside the skyrmion-supporting ROI.

Because the same ROI boundary and the same monitor window are also used for the leakage analysis in Fig.~\ref{fig:results}(d), we define the cumulative leakage up to an additional radial extension $\Delta r_\mathrm{out}$ as
\begin{equation}
  L_\mathrm{cum}(\Delta r_\mathrm{out})
  =
  \frac{\displaystyle\iint_{r_1<r\le r_1+\Delta r_\mathrm{out}}
    \left(|E_x|^2+|E_y|^2+|E_z|^2\right)\,dA}
       {\displaystyle\iint_\Omega \left(|E_x|^2+|E_y|^2+|E_z|^2\right)\,dA},
  \qquad
  0\le \Delta r_\mathrm{out}\le r_\mathrm{m}-r_1,
  \label{eq:cumulative_leakage}
\end{equation}
so that the leakage analysis and the FoM concentration term are evaluated on the same physical monitor and differ only in whether the quantity is reported as energy retained inside $A$ or energy accumulated outside $A$. With $r_\mathrm{m}=3.00~\mu\mathrm{m}$ and $r_1\approx1.55~\mu\mathrm{m}$, the maximum available radial extension within the monitor is $r_\mathrm{m}-r_1\approx1.45~\mu\mathrm{m}$.

To enforce the constituent separation required by the desired field in the annular region between the first zero $r_0$ of $J_0$ and the first zero $r_1$ of $J_1$, $A_\mathrm{ring}=\{(x,y):r_0<r\le r_1\}$, we define the ring-handedness score
\begin{equation}
  H_\mathrm{ring}
  =
  \frac{\displaystyle\iint_{A_\mathrm{ring}} |E_\mathrm{L}|^2\,dA}
       {\displaystyle\iint_{A_\mathrm{ring}} \left(|E_\mathrm{L}|^2+|E_\mathrm{R}|^2\right)\,dA},
  \qquad
  E_\mathrm{R}=\frac{E_x+iE_y}{\sqrt{2}},
  \quad
  E_\mathrm{L}=\frac{E_x-iE_y}{\sqrt{2}},
  \label{eq:ring_handedness}
\end{equation}
where $E_\mathrm{R}$ and $E_\mathrm{L}$ are the circular-basis coefficients obtained by transforming the simulated transverse Cartesian field into the same handedness convention used in Eq.~\ref{eq:desired_field}. This handedness convention is used consistently throughout the manuscript for the polarization-ellipse plots, handedness discussion, and ring-handedness score. In the ideal desired field, the RCP $J_0$ constituent has already vanished throughout $A_\mathrm{ring}$, so any residual RCP content there acts as noise that distorts the intended handedness distribution and perturbs the ROI-integrated skyrmion number. The base figure of merit is then
\begin{equation}
  \mathcal{F}_\mathrm{base}
  = \eta^{\gamma}\,C_\mathrm{ROI}\,H_\mathrm{ring},
  \label{eq:fom_base}
\end{equation}
where $\gamma$ denotes the overlap exponent. The multiplicative form suppresses the score whenever the desired-field overlap is poor, too little total energy is concentrated inside $A$, or the annular handedness is incorrect.

The topological contribution is introduced through the skyrmion density
\begin{equation}
  \rho_\mathrm{sk}(x,y)
  = \mathbf{n}\cdot
  \left(
    \frac{\partial \mathbf{n}}{\partial x}
    \times
    \frac{\partial \mathbf{n}}{\partial y}
  \right),
  \label{eq:skyrmion_density}
\end{equation}
and the corresponding skyrmion number
\begin{equation}
  N_\mathrm{sk}
  = \frac{1}{4\pi}\iint_A \rho_\mathrm{sk}(x,y)\,dx\,dy,
  \label{eq:skyrmion_number}
\end{equation}
where the normalized Stokes vector on the observation plane is
\begin{equation}
  \mathbf{n}(x,y)
  =
  \frac{\left[S_x(x,y),\,S_y(x,y),\,S_z(x,y)\right]}
       {\sqrt{S_x^2(x,y)+S_y^2(x,y)+S_z^2(x,y)}}.
  \label{eq:stokes_vector}
\end{equation}
Numerically, Eqs.~\ref{eq:skyrmion_density} and \ref{eq:skyrmion_number} are evaluated directly on the native Cartesian FDTD grid of the observation plane. The spatial derivatives entering Eq.~\ref{eq:skyrmion_density} are taken from that native discrete Stokes field, and no additional thresholding, smoothing, low-intensity masking, or other post-processing is applied. The reported $N_\mathrm{sk}$ values are therefore obtained exactly by computing $\rho_\mathrm{sk}$ from Eq.~\ref{eq:skyrmion_density} and integrating Eq.~\ref{eq:skyrmion_number} only over the same circular ROI $A$ used in Eq.~\ref{eq:overlap}. No data outside $A$ enter the skyrmion-number calculation; accordingly, the circular field maps shown in Figs.~\ref{fig:optimization}(b,c) and \ref{fig:results}(b,c) visualize the same spatial region used for this topological evaluation.

In the early stages of the continuation schedule, only $\mathcal{F}_\mathrm{base}$ is optimized. Once the projection sharpness reaches $\beta \ge 64$, candidate designs that generate beams with $|N_\mathrm{sk}|$ closer to the ideal unit-skyrmion value $N_{\mathrm{sk},0}=1$ receive an additional reward through
\begin{equation}
  T_\mathrm{sk}
  =
  \exp\!\left[
    -\left(
      \frac{|N_\mathrm{sk}|-N_{\mathrm{sk},0}}
           {\sigma_\mathrm{sk}}
    \right)^2
  \right],
  \label{eq:topology_score}
\end{equation}
so that the total figure of merit becomes
\begin{equation}
  \mathcal{F}
  = \mathcal{F}_\mathrm{base} + w_\mathrm{sk}\,T_\mathrm{sk}.
  \label{eq:fom_total}
\end{equation}
Here $\sigma_\mathrm{sk}$ controls the width of the topological window, and $w_\mathrm{sk}=0$ during the early stages. The detailed stage-dependent schedule for $\gamma$, $w_\mathrm{sk}$, and $\sigma_\mathrm{sk}$ is provided in the Supporting Information.

We perform three-dimensional finite-difference time-domain (FDTD) simulations using the open-source software Meep~\cite{Oskooi2010,Yee1966,Taflove1988}. The forward source used throughout the optimization and final validation is the same normally incident Gaussian beam defined above, launched from the substrate side with beam-waist radius $w_0=1.15~\mu\mathrm{m}$ at the metasurface plane and $45^{\circ}$ linear polarization with respect to the $x$- and $y$-axes. The computational grid has a spatial resolution of 25.0~pixels/$\mu$m, corresponding to a grid spacing of 40.0~nm. Within the design region, the structure is parameterized by a continuous material-density function $h(\mathbf{x}) \in [0,1]$, where $h=1$ and $h=0$ correspond to silicon and air, respectively, through interpolation of the local permittivity. The initial density field is seeded randomly with values between 0.4 and 0.6 to avoid bias toward either material. To ensure manufacturable binary patterns, $h(\mathbf{x})$ is first filtered with a conic kernel~\cite{Bruns2001} and then projected using a subpixel-smoothed projection scheme~\cite{Hammond2025SSP}. The minimum feature size and minimum gap size are both set to 100~nm. For the 0.800~$\mu$m-thick silicon layer considered here, these constraints correspond to a maximum aspect ratio of 8:1, which is compatible with standard nanofabrication processes such as electron-beam lithography and anisotropic dry etching~\cite{Piggott2017,Hammond2021}.

The design variables are updated using the Adam optimizer~\cite{Kingma2015}. To progressively enforce binarization, we employ a continuation schedule for the projection sharpness parameter $\beta$ in the hyperbolic-tangent projection function~\cite{shang2023inverse,kim2025freeform}, with stage values of $\beta = 2$, 4, 6, 8, 16, 32, 64, 128, 256, and 512. The optimized density obtained at each stage initializes the subsequent stage, and the final binary pattern is generated by thresholding the projected density. The complete optimization consists of 255 iterations, each involving one forward and one adjoint FDTD simulation, and requires approximately 45.9~hours of wall-clock time on a dual-socket workstation equipped with two AMD EPYC 9554 processors, providing a total of 128 physical cores (256 hardware threads).

The gradients are evaluated by the adjoint method, which requires only one forward and one adjoint simulation per iteration~\cite{Hughes2018,LalauKeraly2013,Hammond2022,seo2026physics}. For the discretized time-harmonic Maxwell system,
\begin{equation}
  \mathbf{A}\!\left[\varepsilon(\mathbf{r})\right]\mathbf{e}=\mathbf{b},
  \label{eq:maxwell_operator}
\end{equation}
the adjoint field $\mathbf{e}_\mathrm{adj}$ is defined by
\begin{equation}
  \mathbf{A}^{\mathsf T}\mathbf{e}_\mathrm{adj}
  =
  \left(\frac{\partial \mathcal{F}}{\partial \mathbf{e}}\right)^{\mathsf T},
  \label{eq:adjoint_problem}
\end{equation}
where the adjoint source is launched on the same observation plane used for desired-field definition and FoM evaluation, with the source determined by the derivative of the objective with respect to the monitored field components. The gradient with respect to a scalar design parameter $p$ is then
\begin{equation}
  \frac{d\mathcal{F}}{dp}
  =
  -2\,\mathrm{Re}\!\left\{
    \mathbf{e}_\mathrm{adj}^{\mathsf T}
    \frac{\partial \mathbf{A}}{\partial p}
    \mathbf{e}
  \right\},
  \label{eq:adjoint_gradient}
\end{equation}
which reduces to the familiar forward--adjoint field overlap over the design region because the parameter dependence enters only through the permittivity distribution. The sensitivity with respect to the raw density field is obtained by back-propagating this gradient through the interpolation, filtering, and projection operations. Denoting the filtered and projected densities by $h_\mathrm{f}$ and $h_\mathrm{p}$, respectively, the reverse-order Jacobian chain is
\begin{equation}
  \nabla_h \mathcal{F}
  =
  \left(\frac{\partial h_\mathrm{f}}{\partial h}\right)^{\mathsf T}
  \left(\frac{\partial h_\mathrm{p}}{\partial h_\mathrm{f}}\right)^{\mathsf T}
  \left(\frac{\partial \varepsilon}{\partial h_\mathrm{p}}\right)^{\mathsf T}
  \nabla_{\varepsilon}\mathcal{F},
  \label{eq:chain_rule}
\end{equation}
so that the imposed minimum-feature and minimum-gap constraints are incorporated consistently throughout the optimization.

\section{Optimization Trajectory and Device Performance}

\begin{figure*}[!ht]
  \centering
  \includegraphics[width=\textwidth]{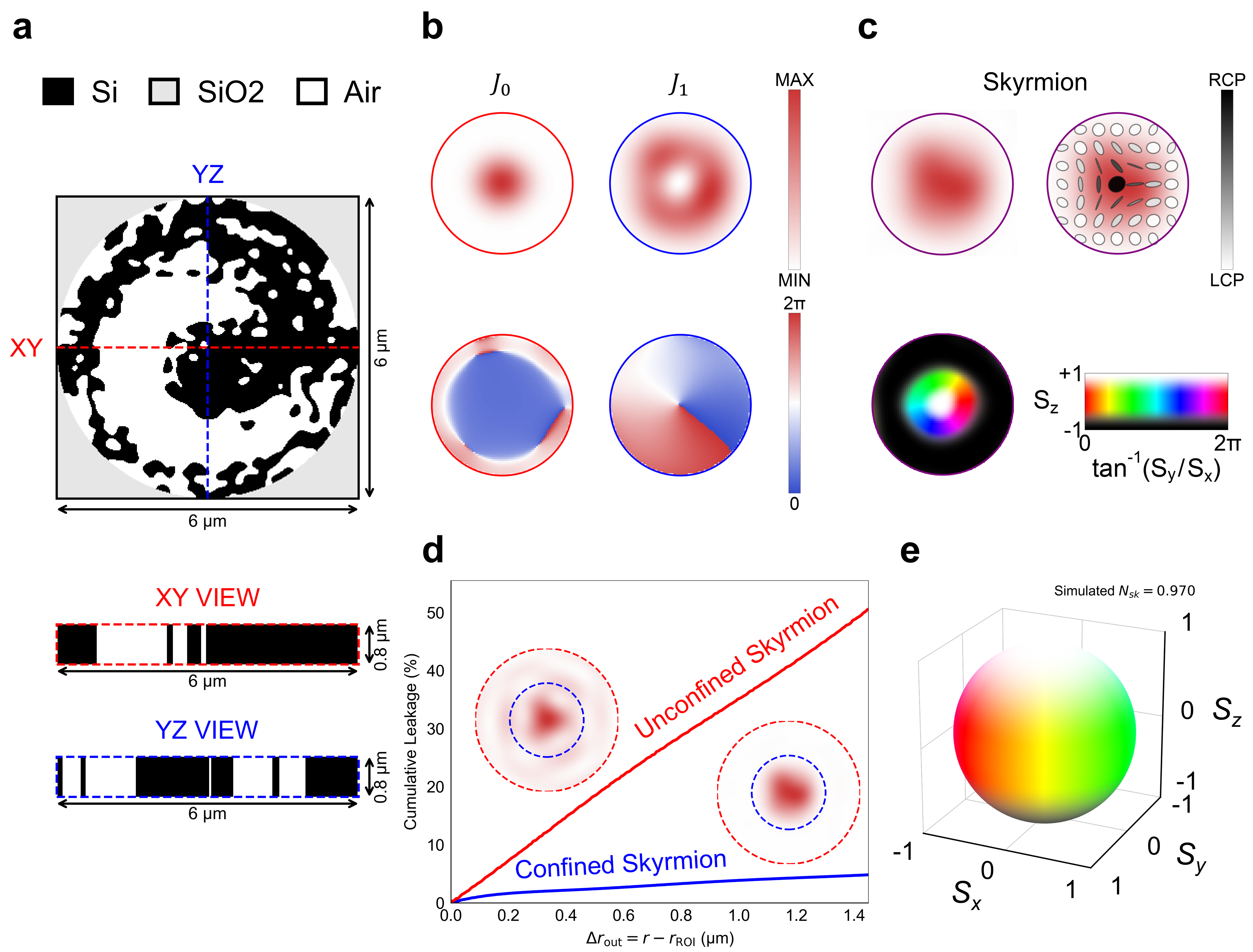}
  \caption{Final optimized metasurface and performance. (a) Binary material layout of the optimized 6.00~$\mu$m-diameter device, together with orthogonal cross-sections through the 0.800~$\mu$m-thick silicon layer on silica. (b) Simulated amplitude (top) and phase (bottom) of the two constituent circular-basis fields at the observation plane, obtained by resolving the simulated transverse field into the same basis used to define the desired right-circularly polarized $J_0$ and left-circularly polarized $J_1e^{i\phi}$ terms; the field maps are shown only within the ROI $A=\{(x,y):r\le r_1\}$. (c) Total skyrmion output within the same ROI, shown as the intensity profile, a polarization-ellipse map plotted on the total-intensity background of the superposed skyrmion field, with ellipse grayscale indicating local handedness, and the skyrmion texture. In the accompanying skyrmion-texture color bar, the vertical axis denotes the normalized out-of-plane Stokes parameter $S_z$, and the horizontal axis denotes the Stokes-space azimuthal angle, written here as $\tan^{-1}(S_y/S_x)$. (d) Cumulative leakage outside the ROI, accumulated from $r=r_1$ outward within the same circular monitor window $\Omega=\{(x,y):r\le 3.00~\mu\mathrm{m}\}$ used in the FoM concentration term, for the inverse-designed confined skyrmion and the corresponding unconfined Bessel-mode skyrmion. (e) Poincar\'{e}-sphere mapping of the simulated Stokes vector over $A$, yielding $N_\mathrm{sk} = 0.970$.}
  \label{fig:results}
\end{figure*}

We first examine the optimization trajectory in Fig.~\ref{fig:optimization}. The FoM in Fig.~\ref{fig:optimization}(a) increases overall in a nonmonotonic but steadily improving manner. Temporary drops occur when the projection-function parameter $\beta$ is increased, as expected in continuation-based topology optimization, because a sharper projection makes the admissible design space more restrictive. The subsequent recovery shows that the structure adapts successfully to progressively stricter binarization and manufacturability constraints. A further gain appears after the skyrmion-number reward is activated at late iterations, indicating that the optimization can be biased toward $|N_\mathrm{sk}| \approx 1$ once a strong observation-plane field match has already been established.

The desired and optimized fields in Fig.~\ref{fig:optimization}(b) are shown after cropping the observation-plane data to the ROI $A=\{(x,y):r\le r_1\}$, because this is the same region used for overlap evaluation and skyrmion-number integration. Within that common region, the optimized field reproduces the compact desired intensity profile and recovers the characteristic skyrmion-density distribution with close visual agreement. Because the skyrmion density is the key quantity from which $N_\mathrm{sk}$ is evaluated, this agreement provides direct evidence that the optimization converges toward the intended topological state.

Representative intermediate states are shown in Fig.~\ref{fig:optimization}(c). These snapshots indicate that different aspects of the desired field converge at different iterations. The compact intensity profile emerges relatively early, whereas the skyrmion-density distribution evolves more gradually and approaches the desired skyrmion-density distribution only at later iterations. This behavior indicates that scalar confinement is easier to establish than the full vectorial topology, which requires coordinated control of amplitude, phase, and polarization across the aperture. At the same time, the grayscale material distribution evolves into a nearly binary silicon--air pattern, confirming that the continuation schedule drives the solution toward a manufacturable geometry.

Figure~\ref{fig:optimization}(d) summarizes the final operating picture: a normally incident $45^{\circ}$-linearly polarized Gaussian beam is transformed by the optimized cylindrical design region into the desired skyrmionic vector field above the metasurface. This schematic therefore connects the optimization variables and observation-plane objective to the final device operation in a compact geometric picture.

We next evaluate the final thresholded device shown in Fig.~\ref{fig:results}. The binary material layout in Fig.~\ref{fig:results}(a) shows that the inverse-design procedure converges to a nontrivial, highly aperiodic silicon--air pattern within the 6.00~$\mu$m-diameter design region while satisfying the imposed thickness and fabrication constraints. As a post-processing step, isolated silicon islands with an area smaller than 0.02~$\mu$m$^2$ are removed from the thresholded design in order to eliminate features that would be difficult to fabricate or mechanically unstable. The cleaned structure is the geometry shown in Fig.~\ref{fig:results}(a), and all performance metrics reported below are evaluated for this final layout. The orthogonal cross-sections further confirm that the optimized structure remains fully binary throughout the 0.800~$\mu$m-thick silicon layer on silica. The absence of any obvious periodic motif indicates that the device response is coordinated across the full aperture rather than synthesized from a repeated unit cell.

Although the thresholded pattern in Fig.~\ref{fig:results}(a) does not admit a simple unit-cell interpretation, its action can still be understood qualitatively in full-wave terms. Because the FoM simultaneously rewards field overlap inside $A$, concentrates total field energy inside $A$ relative to the larger monitor window $\Omega$, and suppresses residual RCP content in the annulus $A_\mathrm{ring}$, the optimizer is driven toward aperture-scale cooperative scattering rather than locally independent phase assignment. The resulting aperiodic silicon--air channels therefore act as coupled radiating elements whose amplitudes and phases are coordinated across the full 6.00~$\mu$m aperture: components that build the central RCP $J_0$-like core and the surrounding LCP $J_1$-like annulus are reinforced coherently within the desired-field region, whereas peripheral components that would otherwise feed the outer Bessel rings are strongly suppressed across the remainder of the monitor. In this sense, the optimized pattern does not remove the constituent-channel physics of the skyrmion construction; instead, it reshapes their collective interference so that the central skyrmion texture is retained while the extended Bessel side lobes are strongly suppressed.

The component-resolved response is shown in Fig.~\ref{fig:results}(b). For clarity, this panel is displayed only within the ROI $A$ and does not show an additional modal projection onto ideal Bessel functions. Instead, it resolves the simulated transverse field on the observation plane into the same circular basis used in Eq.~\ref{eq:desired_field}, so that the plotted quantities are the simulated $E_\mathrm{R}$ and $E_\mathrm{L}$ components themselves. In this representation, the RCP component $E_\mathrm{R}$ is designed to reproduce the truncated $J_0$ constituent primarily in the central region $r\le r_0$ and forms a compact lobe with nearly uniform phase, whereas the LCP component $E_\mathrm{L}$ is designed to reproduce the truncated $J_1e^{i\phi}$ constituent in the surrounding annulus $r_0<r\le r_1$, where the ideal skyrmion construction requires the RCP component to have already vanished. The residual RCP content in that annulus is correspondingly weak, consistent with the role of the ring-handedness term in the FoM. These two constituent fields therefore retain the essential physical roles prescribed in the desired-field construction. Although slight asymmetries remain, as expected for a finite-aperture binary three-dimensional device optimized under full-wave and fabrication-aware constraints, the dominant amplitude and phase characteristics of both components are well preserved. This preservation is consistent with the interpretation above: the aperiodic geometry suppresses unwanted peripheral radiation mainly by coupled redistribution of energy across the aperture rather than by destroying the underlying constituent channels.

The resulting total skyrmion field is shown in Fig.~\ref{fig:results}(c), again cropped to the ROI $A$. The transmitted intensity remains concentrated within the prescribed circular region of interest, and the polarization texture exhibits the expected handedness variation between the central and outer regions. The skyrmion texture is likewise consistent with the desired skyrmionic configuration. In this representation, the accompanying two-dimensional color bar encodes the normalized out-of-plane Stokes parameter $S_z$ vertically and the Stokes-space azimuthal angle, written here as $\tan^{-1}(S_y/S_x)$, horizontally. Quantitatively, the normalized overlap $\eta$ [Eq.~\ref{eq:overlap}] between the simulated and desired vector fields reaches 96.0\% at $\lambda = 1550$~nm, confirming that the observation-plane field synthesized by the metasurface closely matches the desired field.

The principal advantage of the spatial-truncation strategy is quantified in the leakage analysis of Fig.~\ref{fig:results}(d). During optimization, the confinement-control term is the in-ROI energy fraction $C_\mathrm{ROI}$ [Eq.~\ref{eq:roi_concentration}], evaluated over the full circular monitor window $\Omega=\{(x,y):r\le 3.00~\mu\mathrm{m}\}$. Because the ROI boundary is set by the first zero of $J_1$ at $r_1\approx1.55~\mu\mathrm{m}$, the monitor extends well beyond the skyrmion-supporting region. Figure~\ref{fig:results}(d) uses this same ROI boundary and the same monitor window, but reports the complementary information as cumulative outside-ROI leakage $L_\mathrm{cum}(\Delta r_\mathrm{out})$ [Eq.~\ref{eq:cumulative_leakage}] accumulated radially from $r=r_1$ to $r=r_1+\Delta r_\mathrm{out}$. In other words, the FoM rewards concentration of total field energy inside $A$, whereas Fig.~\ref{fig:results}(d) visualizes how much of that same total energy appears outside $A$ as the radial integration range is expanded toward the monitor edge. At the maximum plotted extension $\Delta r_\mathrm{out}=1.40~\mu\mathrm{m}$, the integration radius nearly reaches the monitor boundary at $r=3.00~\mu\mathrm{m}$. For the inverse-designed confined skyrmion, the cumulative leakage based on $|E_x|^2+|E_y|^2+|E_z|^2$ remains below 5.00\% over this range. By contrast, over the same range, the unconfined Bessel-mode superposition reaches approximately 50.0\% cumulative leakage. This order-of-magnitude reduction confirms that the optimized metasurface effectively suppresses the outer Bessel rings that would otherwise broaden the transverse footprint and increase cross-talk between neighboring devices. More broadly, this result provides a quantitative validation of the spatial-truncation strategy introduced in Section~2.1: the optimization is driven not merely to displace peripheral energy outward, but to suppress it substantially within the same fixed monitor used by the FoM.

Finally, the Poincar\'e-sphere mapping in Fig.~\ref{fig:results}(e) spans nearly the full solid angle and yields a simulated skyrmion number [Eq.~\ref{eq:skyrmion_number}] of $N_\mathrm{sk}=0.970$, which remains close to the theoretical desired value of 0.994 shown in Fig.~\ref{fig:desired_field}(e). Here, $N_\mathrm{sk}$ is computed directly over the circular ROI $A$ on the native observation-plane FDTD grid from Eqs.~\ref{eq:skyrmion_density} and \ref{eq:skyrmion_number}, without thresholding, smoothing, or other additional post-processing; no field data outside the $J_1$ first-zero boundary enter this integration. Together with the localized skyrmion-density distribution in Fig.~\ref{fig:optimization}(b), this result indicates that the optimized metasurface preserves the intended polarization topology while substantially reducing the surrounding spatial footprint. Taken together, Fig.~\ref{fig:results} shows that the final binary, fabrication-compatible device simultaneously achieves strong field confinement, high desired-field fidelity, and robust skyrmionic topology.

\section{Conclusion}

By treating the desired skyrmion field itself as the observation-plane quantity to be matched, we inverse-designed a 6.00~$\mu$m-diameter silicon metasurface that generates a spatially isolated optical skyrmion. The proposed framework combines a spatially truncated circularly polarized Bessel-field construction with an observation-plane FoM that directly couples transverse-field overlap inside the ROI, total-energy concentration within that ROI over the full monitor, annular handedness control between the $J_0$ and $J_1$ zeros, and late-stage topological guidance. The resulting aperiodic binary pattern operates through collective aperture-scale interference that preserves the central skyrmion texture while suppressing the outer Bessel side lobes. At $\lambda = 1550$~nm, the optimized metasurface reproduces the prescribed field with a normalized overlap of 96.0\%, limits the cumulative leakage outside the region of interest to below 5.00\%, and yields a simulated skyrmion number of $N_\mathrm{sk}=0.970$.

The distinctive contribution of the present work is the explicit combination of a spatially truncated desired skyrmion field, a fabrication-constrained binary dielectric metasurface, and full-wave adjoint topology optimization, together with quantitative validation of leakage suppression and preserved skyrmion topology. More broadly, the results show that compact skyrmion emitters need not inherit the extended transverse footprint of conventional Bessel superpositions once confinement is built directly into the desired field. The same desired-field inverse-design strategy should be extendable to different skyrmion sizes, desired topological numbers, and related vector-beam objectives. Future work may incorporate broadband or fabrication-robust formulations and validate the present design experimentally through nanofabrication and optical characterization.

\section*{Data availability}
The data and simulation codes that support the findings of this study are available from the corresponding author upon reasonable request.

\bibliography{references_cleaned_verified}

\end{document}